\documentclass[preprint,12pt]{elsarticle}

\usepackage{amssymb}
\usepackage{amsmath}

\usepackage[utf8]{inputenc} 
\usepackage[T1]{fontenc}    
\usepackage{hyperref}       
\usepackage{url}            
\usepackage{booktabs}       
\usepackage{nicefrac}       
\usepackage{microtype}      
\usepackage{xcolor}         
\usepackage{multirow, array, graphicx, subcaption, enumerate, amsfonts}

\newcolumntype{P}[1]{>{\centering\arraybackslash}p{#1}}

\journal{NeuoImage: Reports}

\begin{document}

\begin{frontmatter}



\title{On The Causal Network Of Face-selective Regions In Human Brain During Movie Watching}

 \author[label1]{Ali Bavafa}
 \ead{bavafaali@ut.ac.ir}

 \author[label1]{Gholam-Ali Hossein-Zadeh\corref{cor1}}
 \ead{ghzadeh@ut.ac.ir}
 \cortext[cor1]{Corresponding author}
 \affiliation[label1]{organization={School of Electrical and Computer Engineering, College of Engineering, University Of Tehran},
             addressline={North Kargar st.},
             city={Tehran},
             postcode={1439957131},
             state={Tehran},
			 country={Iran}}

\begin{abstract}
	Understanding the causal interactions in some brain tasks, such as face processing, remains a challenging and ambiguous process for researchers. In this study, we address\- this issue by employing a novel causal discovery method -\textit{Directed Acyclic Graphs via M-matrices for Acyclicity (DAGMA)}- to investigate the causal structure\- of the brain's face-selective network and gain deeper insights into its mechanism. Using fMRI data of natural movie stimuli, we extract causal network of face-selective regions and analyze how frames containing faces influence this network. Specifically, our findings reveal that the presence of faces in the stimuli, causally affects the number of identified connections within the network. Additionally, our results highlight the crucial role of subcortical regions\- in satisfying\- causal sufficiency, emphasizing it's importance in causal studies of brain. This study provides a new perspective on understanding the causal architecture\- of the face-selective network of the brain, motivating further research on neural causality.
\end{abstract}


\end{frontmatter}



\section{Introduction}

With the advancement of Non-invasive brain imaging technologies like functional Magnetic Resonance\- Imaging (fMRI) we can assess brain activity of a living subject. Localization of brain activity in response to a cognitive task as done in conventional fMRI studies doesn’t provide us with enough information to understand brain's mechanism. For that, one must know how brain regions\- are interacting with each other. Therefore, considering brain as a network, we are interested in it's information\- flow. This is the question that brain connectivity models are trying to answer \cite{saetia2021constructing}. The most basic connectome, i.e. structural connectome, represents the physical connections between brain regions, typically based on white matter tracts that are derived from diffusion MRI \cite{sporns2005human}. More advanced types of connectivity are functional and effective connectivity (EC). Functional connectivity (FC) depicts pattern of co-activation between brain regions, representing statistical dependency of neuronal activities. At the apex these models, is effective connectivity which aims to capture causal relationship of the regions through measuring the influence that one neural system exerts over another \cite{smith2013functional, friston2011functional}. As in other causal discovery problems, the goal of EC is the extraction of a precise effective connectome of human brain to be used later for inference. As proposed in \cite{venkatesh2021can}, one can employ the model to recommend intervention targets in neural circuits to correct the desired biases while minimizing the affect to other parts of the network. Gaining even a weak understanding of the reward circuit is motivating clinical interventions for individuals suffering from depression, addiction, obsessive compulsive disorder, obesity, etc.

\begin{table}[t]
	\renewcommand{\arraystretch}{1.25}
	\caption{Comparison of causal methods}
	\label{tab:table_1}
	\centering
	\scalebox{0.8}{
		\begin{tabular}{p{2cm}P{2cm}P{2cm}P{1.8cm}P{2cm}P{2cm}P{2.1cm}}
			\toprule
			& \multicolumn{6}{c}{Feature} \\
			\cmidrule(r){2-7}
			\multirow{3}{*}{\textbf{Method}}  & \textbf{Causal Structure Learning} & \textbf{Causal Direction Learning} & \textbf{\small Causal Inference} & \textbf{Guarantee of Causality} & \textbf{Static/ Dynamic Learning} & \textbf{Cycle Detection} \\
			\midrule
			Granger Causality\- & No & Yes & No & No & Both & Yes \\
			Dynamic Causal Model & No & No & No & No & Dynamic Only & Yes \\
			Structural Equation\- Model & No & No & Yes & Yes & Both & Yes \\
			Probabilistic Graphical Models\- & Yes & Yes & Yes & Yes & Both & No \\
			\bottomrule
		\end{tabular}	}
\end{table}

Existing causal methods can be classified into two categories: data-driven methods and parameter learning or model fitting methods. In the latter category there are two major approaches that has been excluded in this study: Dynamic Causal Model (DCM;\cite{friston2003dynamic}) and structural equation modelling (SEM;\cite{wright1921correlation, mclntosh1994structural}). The primary reason is that neither method is able to effectively search across the full range of possible network topologies. In general, both approaches need (at most) a few potential networks to be hypothesised and compared with their respective modelling approaches \cite{smith2011network}. Instead we will focus on the data-driven methods and especially graphical approaches, that has been shown \cite{spirtes2001causation} to perform more accurate in identifying underlying causal structure especially for large scale network. Table \ref{tab:table_1} Shows a brief, high-level comparison of different methods. Although first attempts on graphical models dates back to even before Granger-causality (GC) \cite{granger1969investigating}, to the early 20th century, recent advances on Directed Graphical Causal models (DCGM) and probablistic view of causality are influenced mostly by J. Pearl’s works (e.g. \cite{pearl2000models}). A conventional method for identifying causal relationships is through interventions or randomized experiments. However, in many cases, this approach is too expensive, too time-consuming, unethical or even impossible. Therefore, revealing causal information by analyzing purely observational data, known as causal discovery, has drawn much attention. Causal discovery algorithms based on DCGMs fall into two categories: (1) constraint-based and (2) score-based. One of the oldest constraint-based algorithms is PC, which identifies the structure of the network by performing conditional independence tests among variables. On the other hand, Greedy Equivalence Search (GES) algorithm is one of the score-based algorithms. While PC starts from full graph and prone it by performing CI tests and some other rules, GES is a two-phase algorithm that starts from empty graph and adds edges to maximize\- a score (first phase) and then removes the edges to improve the score further (second phase). Under certain assumptions and large samples, the two algorithms converge to the same Markov Equivalence Class (MEC). Both algorithms assumes causal sufficiency. This assumption is crucial for their correctness and ability to recover the true causal structure \cite{glymour2019review}. A paradigm shift occured after introduction of \textit{Non-combinatorial Optimization via Trace Exponential and Augmented lagRangian for Structure learning (NOTEARS)} by Zheng et al. \cite{zheng2018dags}, which formulated the learning problem as a continuous constrained optimization task, by leveraging an algebraic characterization of DAGness. \textit{Gradient-based Optimization of dag-penalized Likelihood for learning linear dag Models (GOLEM)} of Ng et al. \cite{ng2020role}, further improved the accuracy of NOTEARS and then the latest method that outperform all previous static structure learning methods is \textit{Directed Acyclic Graphs via M-matrices for Acyclicity (DAGMA)} introduced by Bello et al \cite{bello2022dagma}.

fMRI Studies have shown that the human brain contains several distinct face-selective regions that consistently respond more strongly to face-related stimuli compared to other stimuli \cite{guntupalli2017disentangling, tsantani2021ffa}. Based on their response patterns, faces can be distinguished from other stimuli in regions such as the Fusiform Face Area (FFA), Occipital Face Area (OFA), posterior Superior Temporal Sulcus (pSTS), Anterior Inferior Temporal Lobe (AITL), and several other areas \cite{tsantani2021ffa, rhodes2009fusiform, eick2020occipital}. However, do these regions process the same information? If not, what specific information is encoded in each of these face-selective regions, and how do these regions interact with each other or with other cortical areas? In most studies, the regions involved in face processing are divided into two systems: the “core system” and the “extended system”, which work together \cite{bernstein2018integrated, calder2011oxford}. The FFA, OFA, and pSTS in both hemispheres form the core system, responsible for the initial analysis of faces \cite{rhodes2009fusiform, tsantani2021ffa}. The extended system includes brain regions that can work in coordination with the core system to extract different meanings and higher-level features from observed faces \cite{bernstein2018integrated, calder2011oxford}. Unlike the core system regions, there is less consensus on the exact regions comprising the extended system. Different studies have reported various regions, including the ATL, amygdala, precuneus, mPFC, and anterior parts of the brain \cite{bernstein2018integrated, fox2009correlates}. These regions are involved in higher-level cognitive processes such as semantic processing, emotional evaluation, motivation, and personality trait analysis \cite{bernstein2018integrated, calder2011oxford, fox2009correlates}.

Large size of the whole-brain networks and limited sample size, make the discovery of an accurate EC a challenging task. Using a prior information favoring sparsity of the graph, can significantly increase accuracy. Another challenge is to assess accuracy of extracted EC, while there is no access to the ground truth. Works like \cite{bagheri2023brain}, address these challenges. Studying the brain dynamically, is crucial for understanding its complex mechanism. This is addressed by Dynamic Effective Connectome (DEC) methods like \cite{bagheri2024discovering}. Causal sufficiency assumption, plays a vital role on the validity of any discovered causal structure; A fact that is often overlooked in causal studies of brain networks. In \cite{bagheri2024algorithmic}, an algorithmic identification was approach proposed for determining essential exogenous nodes that satisfy the critical need for causal sufficiency to adhere to it in such inquiries.

\textbf{Contributions:} In this study, we provide a performance comparison of GC, as one of the most widely used causal methods, and two recent gaphical methods namely GOLEM and DAGMA using synthetic fMRI data and we show that DAGMA outperforms other methods. Then we apply DAGMA on real fMRI data from movie watching task, provided by Human Connectome Project (HCP). Concretely, our contributions are as follows:

\begin{itemize}
	\item We extract the causal network of face-selective network of a group of 30 subjects during watching natural movie stimuli.
	\item We show that the presence of movie frames containing face have causal effect on the number\- of the identified causal connections in the network.
	\item We show the essential role of inclusion of subcortical regions in the discovered causal structure as a confounding node of the network.
\end{itemize}

	\section{Method}
In this section, we begin by reviewing the relevant material from the literature. Specifically, we discuss Granger causality, Bayesian Networks, and two causal structure learning methods: GOLEM and DAGMA. Then the concept of Average Treatment Effect (ATE) is introduced to aid our causal inference. Next, in the data subsection, we describe the process of generating synthetic data and present the HCP datasets used in our analyses, providing details about data acquisition and preprocessing procedures.

\subsection{Preliminaries}

\subsubsection{Granger Causality (GC)}
For two stationary timeseries, $X$ ``Granger-causes'' $Y$ if considering past values of $X$ yields more accurate prediction
of future of $Y$ compared to only considering past values of $Y$. In other words, there is some unique information in X relevant for Y that is not contained in Y's past as well as the past of ``all the information in the universe''. In practice however, typically only Y's past is used (bivariate Granger causality). Among different variations of Granger causality, we consider the most typical linear Auto-Regressive (AR) model:

\begin{equation}
	\mathbf{X}_t=
	\sum_{\tau=1}^{\tau_{max}}
	\Phi(\tau)	\mathbf{X}_{t-\tau} + \eta_t
	\label{eq:eq1}
\end{equation}

where, $ \mathbf{X}_t = (X_t^1, \dots, X_t^N) $, $ \Phi(\tau) $ is the $N \times N$ coefficient matrix at lag $ \tau $, $ \tau_{max} $ some maximum time lag, and $ \eta $ denotes an independent noise term. In the equation \eqref{eq:eq1}, $X^i$ Granger-causes $X^j$ if any of the coefficients $ \Phi_{ji}(\tau) $ at lag $ \tau $ is non-zero. A non-zero $ \Phi_{ji}(\tau) $ can then be denoted as a causal link $X_{t-\tau}^i \rightarrow X_t^j$ at lag $\tau$. Another option is to compare the residual variances of the VAR fitted with and without including the variable $ X^i $ \cite{runge2018causal}.

\subsubsection{Bayesian Networks (BNs)}

BNs are a class of Probabilistic Graphical Models (PGMs) that consider the underlying relationship among variables\- as a Directed Acyclic Graph (DAG). The term “Bayesian” is used due to the use of Bayesian decomposition rule, i.e. chain rule which simplifies the probability distribution, otherwise there is nothing Bayesian about it. Specifically, for $N$ nodes in the network, $X_1, \dots, X_{N_G}$, the joint probability distribution. is expressed as:

\begin{eqnarray}
	p(\mathbf{X}_{1:N_G})  &=&  p(X_1)p(X_2|X_1)p(X_3|X_2,X_1)\dots p(X_{N_G}|X_1, \dots, X_{N_G - 1}) \nonumber \\ 
	&=& \prod_{i=1}^{N_G}  p(X_i | \mathbf{PA_i})
	\label{eq:eq2}
\end{eqnarray}

Where $\mathbf{PA_i} = \{X_j: X_j \rightarrow X_i \in G\}$ denotes the set of parents, or direct causes, of $X_i$ in $G$. While many other entangled factorizations are possible, only \eqref{eq:eq2} decomposes the joint distribution into causal conditionals, or causal mechanisms, $p(X_i |\mathbf{PA_i})$, which can have a meaningful physical interpretation (Causal Markov Condition). A BN has: (1) a set of random variables as it’s nodes, (2) a set of directed edges, where lack of edge between two nodes implies Conditional Independence (CI), and (3) a joint probability distribution over the possible values of all of the variables. For a Bayesian Network to be Causal Bayesian Network (CBN), directed edges $X_j \rightarrow X_i$ must represent a direct causal effect of $X_j$ on $X_i$ which enables reasoning about the outcome of interventions using the \textit{do-operator}.

Sometimes, we can prune the edges of a CBN using domain knowledge (or other information like temporal ordering) to uncover the causal relationships between variables When such knowledge is unavailable or incomplete, we need to perform \textit{causal discovery}, i.e. causal structure learning from available data. This necessitates \textit{faithfulness} condition \cite{glymour2019review, mumford2014bayesian, scholkopf2022statistical}. Another crucial assumption needed to make sure the true causal structure is being identified is \textit{causal sufficiency} assumption as presented in \cite{spirtes2001causation} which is implicitly considered in the following causal discovery algorithms.

\subsubsection{Causal structure learning}

\paragraph{\textbf{GOLEM}}

Let's denote a DAG model defined over a set of random variables $X = (X_1, \dots, X_{d})$, by \mbox{$G = (V (G), E(G))$}  and the joint distribution $P(X)$ (with density $p(x)$). GOLEM assumes linear\- DAG model with Gaussian noise. In terms of linear Structural Equation Models (SEMs), it can be represented by $X_i = B_i^T X + N_i$ equation for each of the variables, where $B_i$ is a coefficient vector\- and $N_i$ is the exogenous noise variable corresponding to variable $X_i$. In matrix form, if follows $X = B^T X + N$, where $B = [B_1 | \dots | B_d]$ is a weighted adjacency matrix and $N = (N_1, \dots , N_d)$ is a noise vector\- with independent elements. The structure of $G$ is defined by the nonzero coefficients\- in $B$, i.e., $X_j \rightarrow X_i \in E(G)$ if and only if the coefficient  in $B_i$ corresponding to $X_j$ is nonzero. Given i.i.d. samples $\mathbf{x}= \{ x^{(k)}\}_{k=1}^n$, the goal is to infer the matrix $B$. The \textit{unconstrained} optimization\- problem that GOLEM considers for it's score function with soft $\ell_1$ penalty and DAG constraints, is as follows\-:

\begin{equation}
	\min _{B \in \mathbb{R}^{d \times d}} \mathcal{S}_i(B ; \mathbf{x})=\mathcal{L}_i(B ; \mathbf{x})+\lambda_1\|B\|_1+\lambda_2 h(B)
\end{equation}

where $i = 1, 2, \lambda_1$ and $\lambda_2$ are the sparsity and DAGness penalty coefficients, respectively. $\|B\|_1$  is defined element-wise, and $h(B) = \operatorname{tr}\left(e^{B \circ B}\right)-d$ is the characterization of DAGness proposed by Zheng et al. \cite{zheng2018dags} in which $\circ$ is the Hadamard or element-wise product. The score functions $\mathcal{S}_i(B ; \mathbf{x}), i =1,2$ correspond to the likelihood objectives for nonequal and equal noise variances, shown in equations \eqref{eq:NV} and \eqref{eq:EV}, respectively \cite{ng2020role}.

\begin{equation}
	\mathcal{L}_1(B ; \mathbf{x})=\frac{1}{2} \sum_{i=1}^d \log \left(\sum_{k=1}^n\left(x_i^{(k)}-B_i^{\boldsymbol{\top}} x^{(k)}\right)^2\right)-\log |\operatorname{det}(I-B)|
	\label{eq:NV}
\end{equation}

\begin{equation}
	\mathcal{L}_2(B ; \mathbf{x})=\frac{d}{2} \log \left(\sum_{i=1}^d \sum_{k=1}^n\left(x_i^{(k)}-B_i^{\boldsymbol{\top}} x^{(k)}\right)^2\right)-\log |\operatorname{det}(I-B)|
	\label{eq:EV}
\end{equation}

\paragraph{\textbf{DAGMA}}
Here the problem is approached from a more generic perspective, considering SEM equations of the form:

\begin{equation}
	X_j = f_j(X, Z_j), \quad \forall j \in \{ 1 \dots d\}
	\label{dagma_SEM}
\end{equation}

where $X = (X_1, \dots, X_d)$ is a d-dimensional random vector, $f_j: \mathbb{R}^{d+1} \rightarrow \mathbb{R}$ is a nonlinear nonparametric\- function in general, and $Z_j$ is an exogenous variable. It considers the Markovian model, which assumes that each $Z_j$ is an independent random variable. Note that each $f_j$ depends only on a subset of $X$ (i.e., the parents of $X_j$) and $Z_j$. In practice, $f$ is replaced with a flexible family of parametrized functions such as deep neural networks to shrink the problem space to finite dimensions. The score function to assess given data is defined as: 

\begin{equation}
	Q(f; X) = \sum_{j=1}^{d} \text{\textbf{loss}}(x_j, f_j(X))
\end{equation}

where \textbf{loss} can be any loss function such as least squares loss or the log-likelihood function. The \textit{constrained} optimization that DAGMA solves is:

\begin{eqnarray}
	& \hspace{0.5em} \min_{\theta} Q(f_{\theta};\mathbf{X}) +\beta_1\|\theta\|_1  \nonumber \\
	& \text{subject to} \hspace{0.5em} h_{ldet}^s(W(\theta)) = 0
\end{eqnarray}

where $Q$ is score function, $f_{\theta}$ denote a model with parameters $\theta$ for the functions $f_j$ in \eqref{dagma_SEM}, e.g., neural networks and $h_{\scriptscriptstyle \text{ldet}}^s(W)$ is the new log-determinant acyclicity characterization introduced, which is defined as:

\begin{equation}
	h_{\text{ldet}}^s(W) \stackrel{\text{def}}{=}-\log \operatorname{det}(s I - W \circ W) + d \log s
\end{equation}

The score function is also augmented with a $\ell_1$ regularizer to promote sparse solutions \cite{bello2022dagma}.

\subsubsection{Causal Inference}
\paragraph{\textbf{Average Treatment Effect (ATE)}}
\label{ATE}
In the broadest form, ATE is defined as the expected difference in potential outcomes when the entire population is hypothetically assigned to treatment versus control. Formally:

\begin{eqnarray}
	ATE &=& \mathbb{E}[Y^1 - Y^0] \nonumber \\
	&=& \mathbb{E}[Y|do(X=1)] - \mathbb{E}[Y|do(X=0)] 
\end{eqnarray}

where $Y^1$ is the potential outcome if all individuals receive treatment $(X=1)$ and $Y^0$ is the potential outcome if all individuals receive control $(X=0)$ \cite{Morgan_Winship_2015}. For binary $Y$, ATE is reduced to:

\begin{equation}
	ATE = \mathbb{P}(Y=1|do(X=1)) - \mathbb{P}(Y=1|do(X=0))
\end{equation}

\subsection{Data}
\subsubsection{Synthetic fMRI data}
A total of 50 datasets generated as follows: Erdos-Renyi graphs of 5 nodes, 5 edges and 500 time points\- where created for each node (procedure similar to \cite{bello2022dagma}). Each time series is then convolved with \verb+spm_hrf+ of \href{https://www.fil.ion.ucl.ac.uk/spm/software/spm12/}{SPM} package which is an implementation of canonical haemodynanmic response function\- (HRF). Gaussian noise $\mathcal{N}(0, 0.8)$ has been added to the resultant signal from previous step. Finally\- it is sampled and used for algorithms.

\subsubsection{Real fMRI data}	
Two real task fMRI dataset has been used in this study: (1) Working memory task \cite{barch2013function} (2) Movie-watching task \cite{finn2021movie}, both of them released by  \href{https://www.humanconnectome.org/}{Human Connectome Project (HCP)}. Both of the datasets has been preprocessed using minimal pipeline provided by HCP, which includes three structural\- pipelines, two functional pipelines and a Diffusion pre\-processing pipeline \cite{glasser2013minimal}. A total of 175 subjects have the data for both of these tasks.

\subsubsection*{Working memory task}
Working memory is assessed using an N-back task in which participants are asked to monitor sequentially\- presented pictures. Participants are presented with blocks of trials that consisted pictures\- from 4 categories: faces, places, tools, and body parts. Subjects performed two runs of the working memory task. Each run contained eight task blocks (25 s each) and four fixation blocks (15 s each). The four different stimulus types (faces, places, tools, and body parts) were presented in separate task blocks. Each task block contained ten trials. On each trial, the stimulus was presented for 2 s, followed by a 500 ms inter-trial interval. Within each run, four blocks used a 2-back working memory task (respond “target” whenever the current stimulus was the same as the one 2-back) and the other four blocks used a 0-back working memory task (respond “target” whenever the current stimulus was the same as the target stimulus presented at the start of the block). A 2.5 s cue indicated the task type (and target for 0-back) at the start of the block. In each block, there were two targets and 2–3 nontarget stimuli (repeated items in the wrong n-back position, either 1-back or 3-back). 

This dataset has been used in this study as ``functional localizer'' to obtain category-specific representations\- \cite{drobyshevsky2006rapid}. Category-selective maps were obtained by statistical comparison of the activation\- for one category versus the average activation for the other three categories. For a given contrast, the maps were based on z-statistic. In this study, face-selective maps was used. Maps from all subjects were projected onto a standard grayordinates space \cite{glasser2016human}. The grayordinates space contained 91,282 cortical and subcortical gray matter voxels/vertices.

\subsubsection*{Movie-watching task}

The fMRI data from movie watching task of HCP, was collected from subjects during watching naturalistic stimuli. There were four total scan sessions acquired over two or three days; we focus here on the first and last session (which we refer to as session 1 and session 2), since these contained the movie-watching runs. REST and MOVIE runs were collected using the same gradient-echo-planar imaging (EPI) sequence. The direction of phase encoding alternated between posterior-to-anterior (PA; REST1, MOVIE2, MOVIE3) and anterior-to-posterior (AP; REST4, MOVIE1, MOVIE4). During\- REST runs, subjects were instructed to keep their eyes open and maintain relaxed fixation on a projected bright crosshair on a dark background. Within a session, REST runs were always\- acquired first, followed by the movie runs in a fixed order, such that session 1 consisted of REST1, MOVIE1, and MOVIE3, and session 2 consisted of REST4, MOVIE2, and MOVIE4.
During MOVIE runs, subjects passively viewed a series of video clips with audiovisual content. Each MOVIE run consisted of 4 or 5 clips, separated by 20s of rest (indicated by the word “REST” in white text on a black background). Two of the runs, MOVIE1 and MOVIE3, contained clips from independent films (both fiction and documentary) made freely available under Creative Commons license on Vimeo. The other two runs, MOVIE2 and MOVIE4, contained clips from Hollywood films. The last clip was always a montage of brief (1.5 s) videos that was identical across each of the four runs (to facilitate test-retest and/or validation analyses). Each REST run was 900 TRs, or 15:00 min, in length. MOVIE runs 1-4 were 921, 918, 915, and 901 TRs, respectively.

\begin{figure}[ht]
	\begin{center}
		\includegraphics[width=0.8\textwidth]{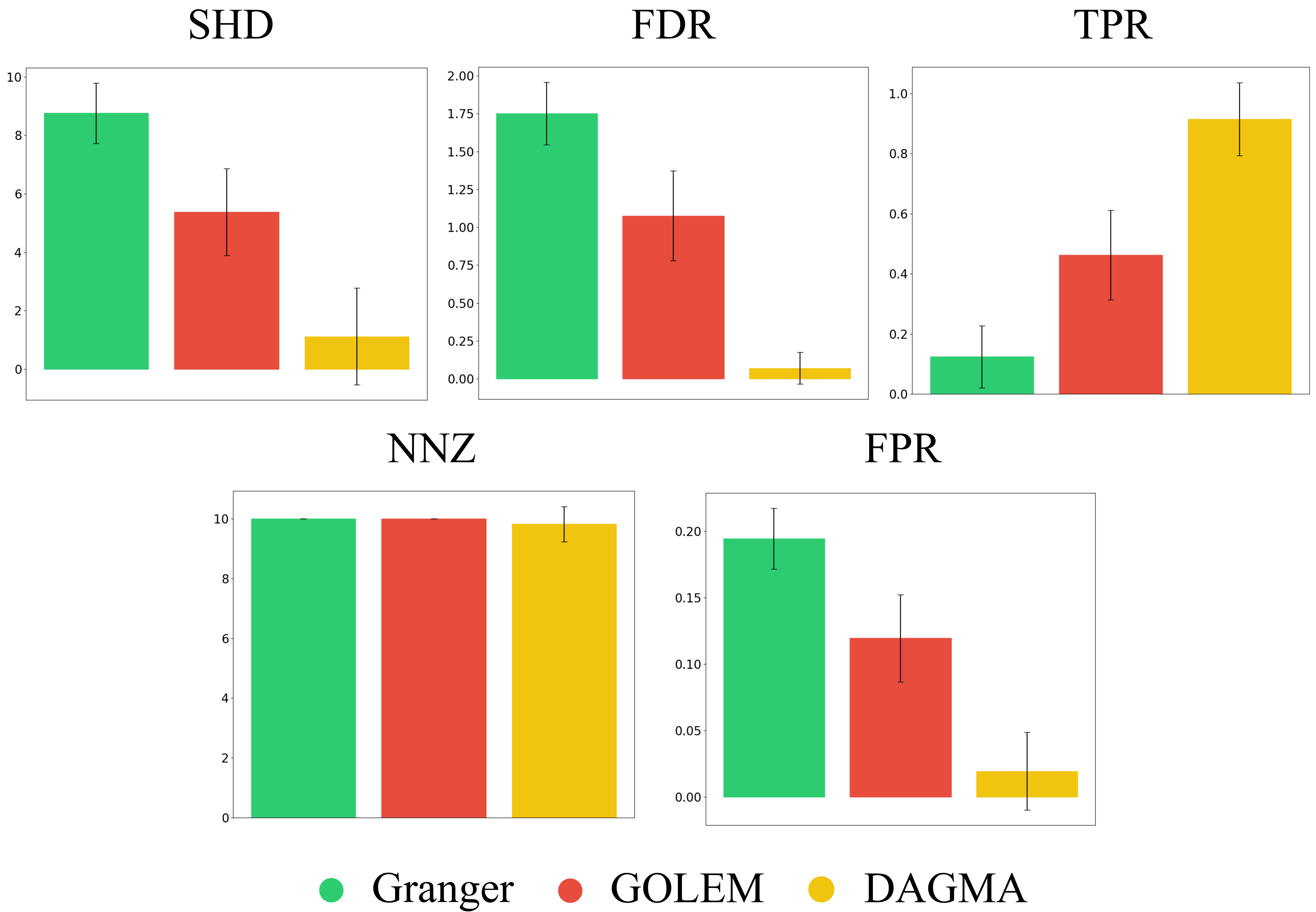}
		\caption{Comparison of performance of data-driven methods across 50 synthetic datasets in terms of mean and standard deviation across multiple metrics: Structural Hamming Distance (SHD), False Discovery Rate (FDR), True Positive Rate (TPR), False Positive Rate (FPR) and Number of Non-Zero elements (NNZ).}
		\label{fig:synthetic}
	\end{center}
\end{figure}

\begin{table}[htbp]
	\renewcommand{\arraystretch}{1.25}
	\caption{Performance (mean $\pm$ standard deviation) of data-driven methods across 50 synthetic fMRI data}
	\label{tab:synthetic}
	\centering
	\scalebox{0.85}{
		\begin{tabular}{cccc} 
			\toprule
			& \multicolumn{3}{c}{Method} \\
			\cmidrule(r){2-4}
			Metric & Granger & GOLEM & DAGMA \\
			\midrule
			Structural Hamming Distance (SHD) & $8.76 \pm 1.030$ & $5.38 \pm 1.481$ & $1.12 \pm 1.657$ \\
			False Discovery Rate (FDR) & $1.752 \pm 0.206$ & $1.076 \pm 0.296$ & $0.0699 \pm 0.104$ \\
			True Positive Rate (TPR) & $0.124 \pm 0.103$ & $0.462 \pm 0.148$ & $0.914 \pm 0.121$ \\
			False Positive Rate (FPR) & $0.194 \pm 0.023$ & $0.119 \pm 0.033$ & $0.019 \pm 0.029$ \\
			\bottomrule
		\end{tabular}}
\end{table}

	\section{Results}

\subsection{Results of synthetic data}	

We executed Granger, GOLEM and DAGMA methods on 50 synthetic fMRI datasets, thresholded the connectivity matrices and kept 10 edges having the largest weights and finally binarized the result. Figure \ref{fig:synthetic} and table \ref{tab:synthetic} show the results in terms of mean and standard deviation across multiple metrics: Structural Hamming Distance (SHD), False Discovery Rate (FDR), True Positive Rate (TPR), False Positive Rate (FPR) and Number of Non-Zero elements (NNZ). For definition of these metrics see \nameref{appendix}.

As shown in the figure \ref{fig:synthetic} and table \ref{tab:synthetic}, DAGMA outperforms other methods in all metrics. NNZ metric shows that all of the methods have found similar number of non-zero elements. However, TPR and FDR metrics show that DAGMA has found more correct edges and has lower error. TPR and FDR metrics gain more value as network grows in size where the ground truth is usually a sparse matrix. \cite{bello2022dagma} shows that DAGMA's superiority to other methods in the large networks maintained or improved. This is of significant importance, namely in the whole-scale causal studies of the brain. Also note that Granger has perform the worst, in comparison to graphical methods. Setting performance matrics aside, Granger also raises some other concerns, e.g. in the case of brain study with fMRI data, neuronal activity occurring first in one area and then in the second area may be seen in the haemodynamic responses in the second area, before the first. This violates the temporal precedence assumption in the GC which is taken as a measure of causality \cite{friston2009causal}. Furthermore, it is a prediction framework. It can't answer causal questions related to consequences of interventions and counterfactuals \cite{biswas2022statistical}.

\subsection{Results of real data}

\subsubsection{Face-selective network}
\begin{figure}[htbp]
	\begin{center}
		\includegraphics[width=0.7\textwidth]{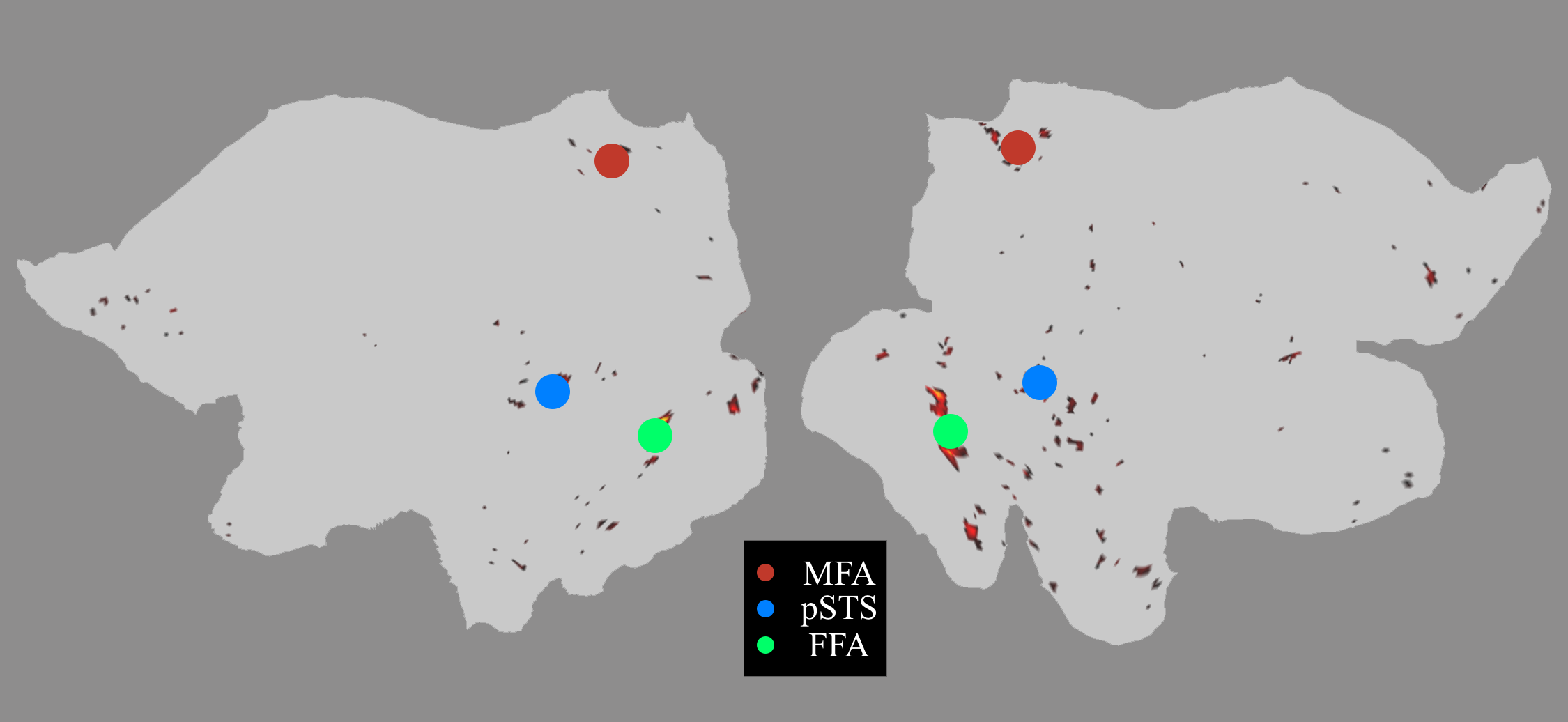}
		\caption{Top 1 \% cortical vertices, i.e. face-selective vertices on the 2D flat map of left and right hemispheres and the center of selected clusters shown in solid-color circles.}
		\label{fig:cluster_center}
	\end{center}
\end{figure}

A subset of 30 subjects was selected for this study. First, face-selective maps were averaged across the subjects. Next, face-selective vertices/voxels were defined as the top 1\% of voxels (913 out of 91,282) with the highest z-values in the given contrast. The 99th percentile corresponded to a cutoff z-value of 4.354. Based on remained cortical vertices, six clusters were selected with reference to the \cite{abbasi2020genetic} study based on the size of active clusters(figure \ref{fig:cluster_center}). Table \ref{tab:coordinate} lists their corresponding center coordinates. Of the total 913 vertices/voxels, 238 were located in the subcortical area, which was considered as another cluster to ensure the validity of the causal sufficiency assumption \cite{spirtes2001causation}. Together, these clusters formed a total of seven nodes for our network after averaging their corresponding time series.

\subsubsection{Movie clips}
\begin{figure}[htbp]
	\centering
	\begin{subfigure}[b]{0.49\textwidth}
		\centering
		\includegraphics[width=\textwidth]{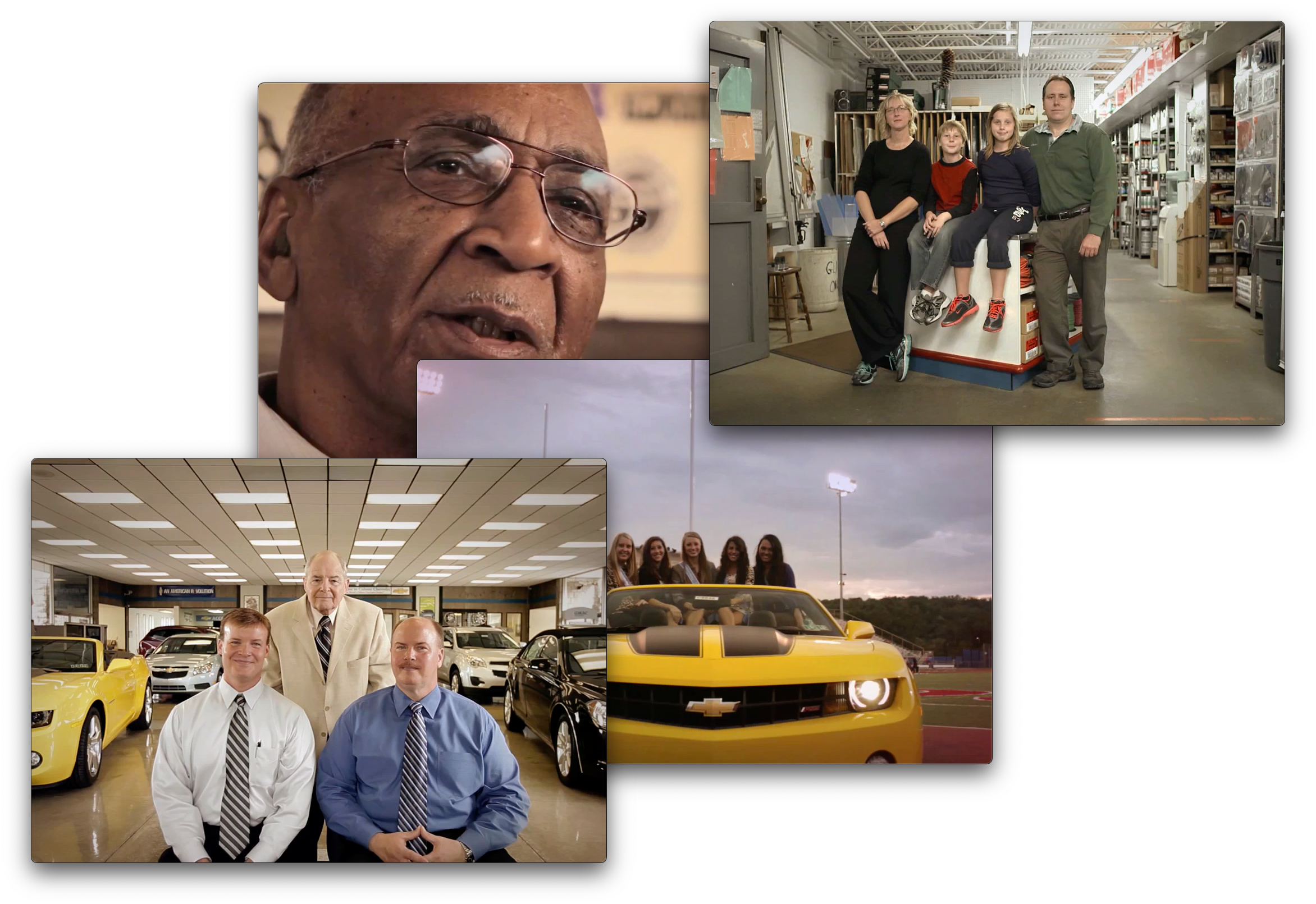}
		\caption{}
		\label{fig:bridge_frame}
	\end{subfigure}
	\hfill
	\begin{subfigure}[b]{0.49\textwidth}
		\centering
		\includegraphics[width=\textwidth]{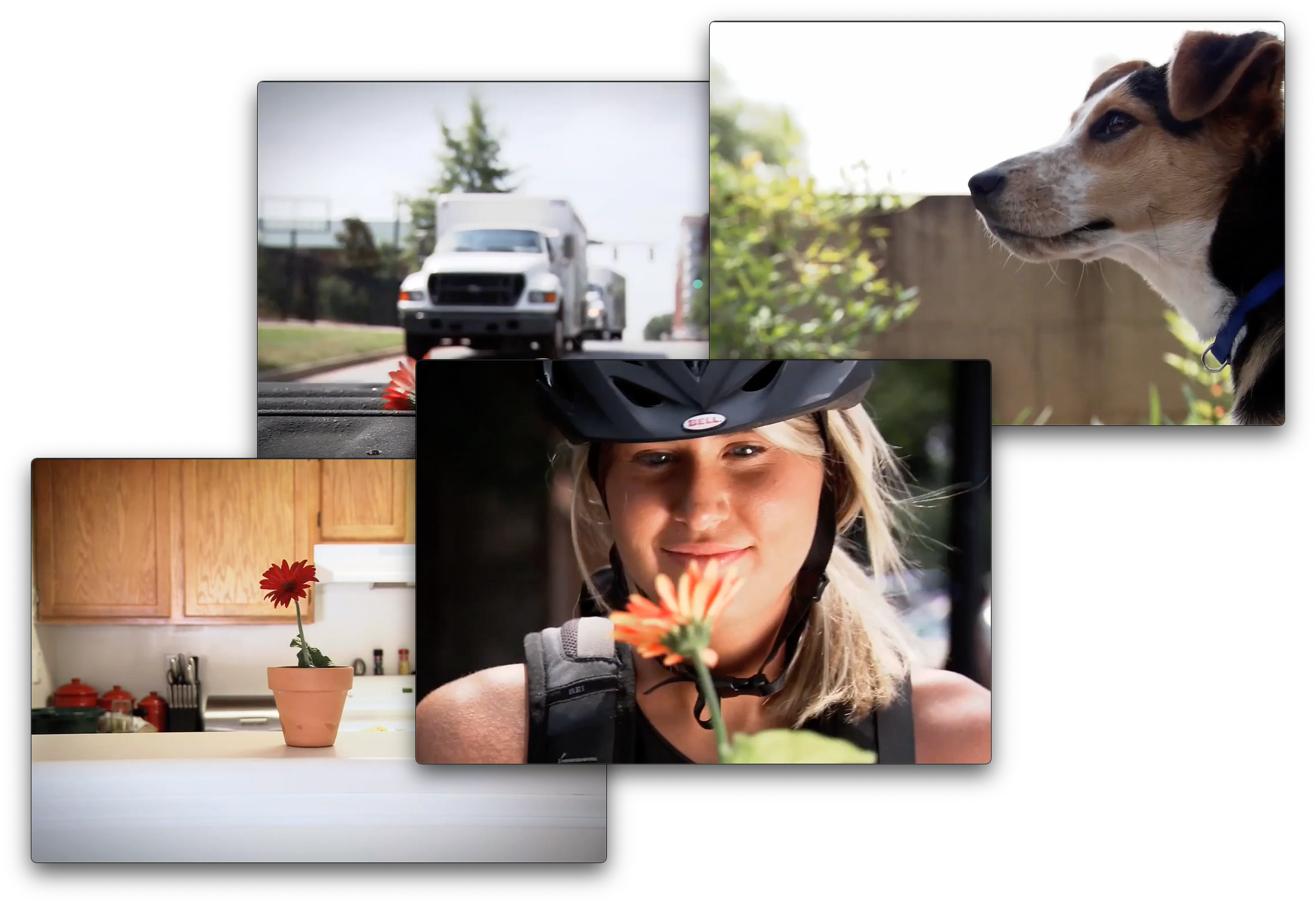}
		\caption{}
		\label{fig:flower_frame}
	\end{subfigure}
	\caption{A few frames from the (a) Bridgeville (b) Flower movie clips.}
	\label{fig:frames}
\end{figure}

Table \ref{tab:clip_details} shows the details of selecteed movie clips for this study. To quantify the number of movie frames containing faces in each clip, the following steps were taken: Images were extracted from each movie clip at a frequency of 1 Hz using the \href{https://www.ffmpeg.org}{ffmpeg} package (originally, movie files contained 24 frames a second). \verb+EfficientNet_B3+ has been finetuned on a small subset of frames to classify them into containing/not containing faces. Then, used the finetuned network to classify all of the images. The result has been double checked by human reviewer. Figure \ref{fig:frames} shows a few frames from contents of each clip. The ratio of number of face frames to all frames for Bridgeville movie clip was $58 \%$ and for Flower  movie clip was $ 1 \%$. 

\begin{figure}[htbp]
	\centering
	\begin{subfigure}[b]{\textwidth}
		\centering
		\includegraphics[width=\textwidth]{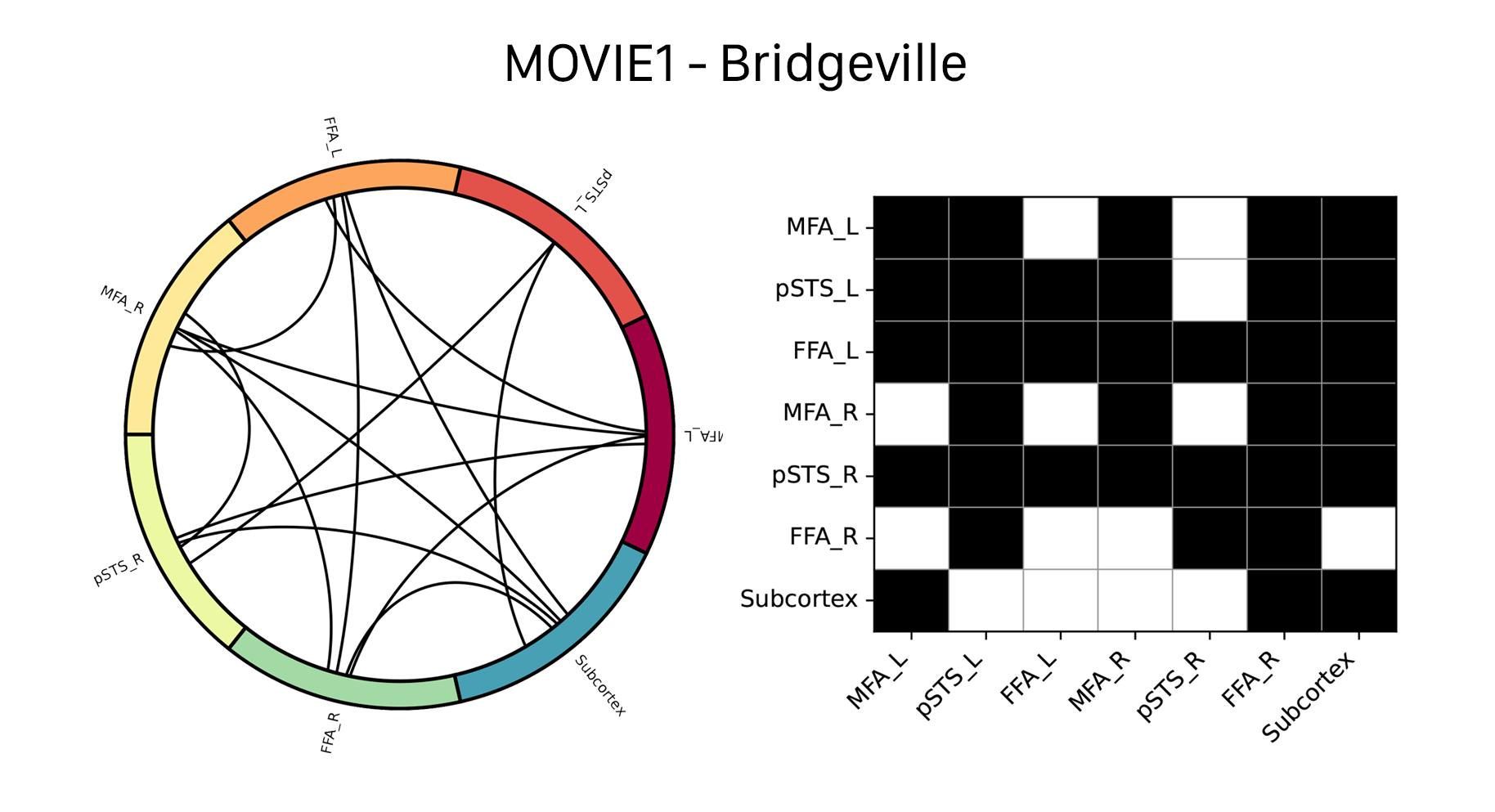}
		\caption{}
		\label{fig:bridge_connectome}
	\end{subfigure}
	\hfill
	\begin{subfigure}[b]{\textwidth}
		\centering
		\includegraphics[width=\textwidth]{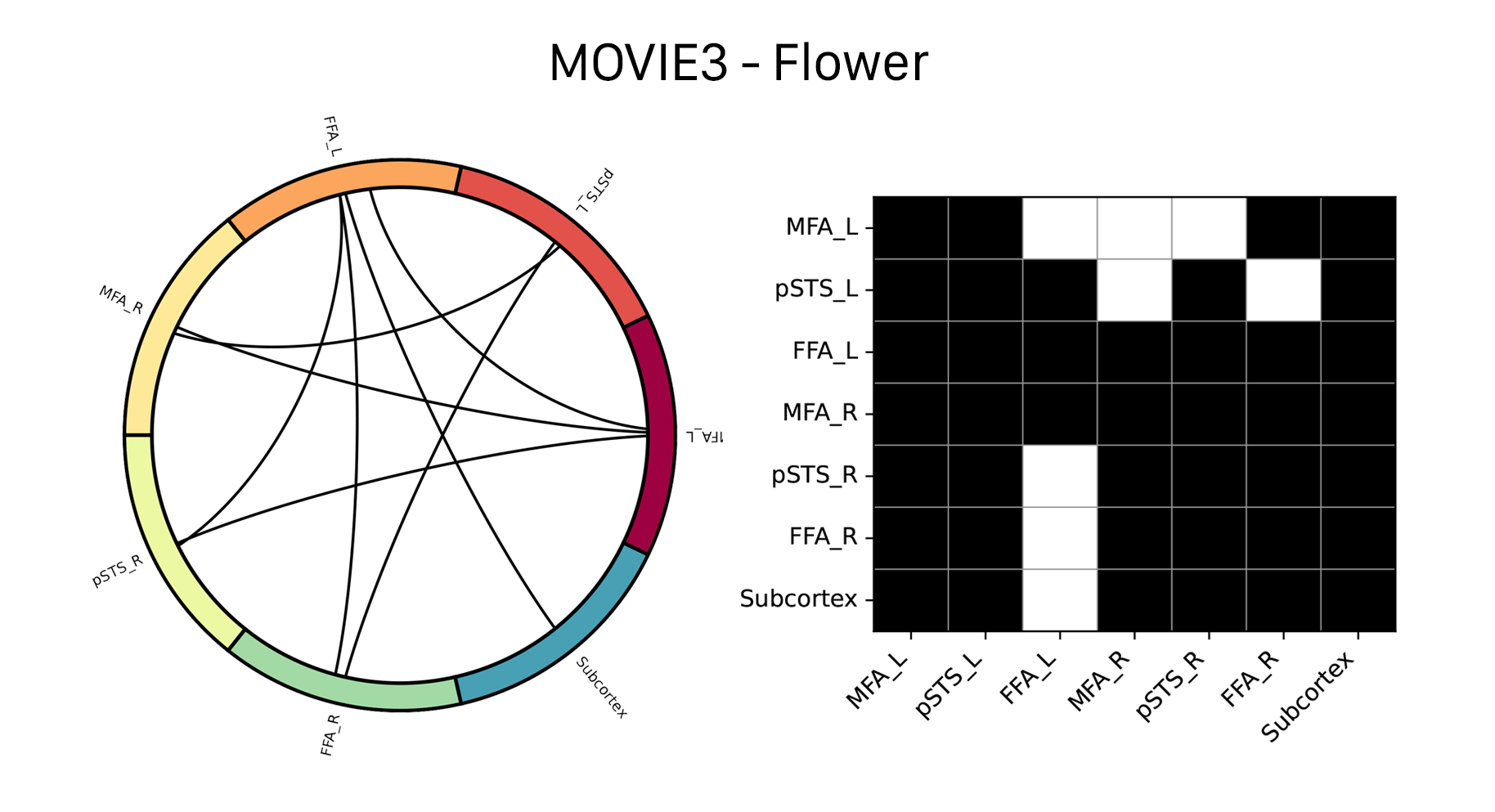}
		\caption{}
		\label{fig:flower_connectome}
	\end{subfigure}
	\caption{Connectogram on the left showing causal connections in undirected symmetric form (presence or absence of a causal connection) and binary connectivity matrix on the right showing directed causal connections of the face-selective network for (a) Bridgeville (b) Flower movie clips. In the connectivity matrix displayed on the right, each row represents a cause and each column represents an effect. White squares indicate the presence of a connection, while black squares indicate its absence.}
	\label{fig:connectome}
\end{figure}

The causal network of face-selective network during watching aformentioned clips have been extracted\- using DAGMA, thresholded and binarized. Figure \ref{fig:bridge_connectome} and \ref{fig:flower_connectome} shows the causal connections for the Bridgeville and Flower movie clips in the form of directed and undirect ECs, respectively. Connectogram on the left shows causal connections in the undirected form, where the connectome is primarily determined by the presence or absence of an edge, leading to a symmetric connectome.  Binary connectivity matrix on the right shows directed EC, where both the existence or absence of edges and the direction of edges are discovered, leading to an asymmetric connectome. As can be seen in figure \ref{fig:connectome}, Bridgeville clip has 14 edges whereas this number for Flower clip decreases to 8. This means more regions of the Bridgeville network are interacting with each other and they are more active, compared to the Flower's network.

\begin{table}[ht]
	\renewcommand{\arraystretch}{1.25}
	\caption{Details of the selected movie clips \cite{finn2021movie}}
	\label{tab:clip_details}
	\centering
	\begin{tabular}{ccc>{\centering\arraybackslash}p{5.3cm}}
		\toprule
		Run & Clip name & Duration(min:sec) & Description\\
		\midrule
		\multirow{2}{*}{MOVIE1} & \multirow{2}{*}{Bridgeville} & \multirow{2}{*}{03:41} & {\scriptsize People describe why they love living in small-town America; a collection of scenes from community life.}\\
		\multirow{2}{*}{MOVIE3} & \multirow{2}{*}{Flower} & \multirow{2}{*}{03:00} & {\scriptsize A flower escapes its pot and goes on a journey through the neighborhood.} \\
		\bottomrule
	\end{tabular}
\end{table}

\subsubsection{Causal effect analysis}
Let $X$ denote whether subjects were exposed to face-containing content, and $Y$ indicate the presence of a causal edge in the network. Since the only difference between stimuli lies in the clip content, and our analysis is restricted to the network of face-selective regions—thereby isolating the effect of face frames— according to ATE (see \ref{ATE}), we conclude that the presence of face frames, have causal effect on the identified number of edges. To quantify this effect, we let $\mathbf{Y} = (Y_{12}, \dots, Y_{ij}, \dots,Y_{d-1d})^T$, where $d$ is the number of nodes in the graph. $Y_{ij}, 1 \le i,j \le d, i \ne j$ is a random variable sampled from Bernouli distribution ($Y_{ij} \sim \mathcal{B}(p)$) modelling the probability, for existence of an edge between two arbitrary nodes, $i$ and $j$. For each network, $p$ is considered as edge density defined as the number of existing edges divided by the number of possible edges \cite{wang2009parcellation}. A DAG with d=7 nodes, thus would have a total of $d(d-1)/2 = 21$ possible edges. Assuming edges as being independent and identically distributed (i.i.d), we have:

\begin{eqnarray}
	ATE &=& \mathbb{P}(\mathbf{Y}=\mathbf{1_d}|do(X=1)) - \mathbb{P}(\mathbf{Y}=\mathbf{1_d}|do(X=0)) \nonumber \\
	&\overset{\textit{i.i.d.}}{\implies}& \prod_{k} \mathbb{P}(Y_{k}=1|do(X=1)) - \prod_{k} \mathbb{P}(Y_{k}=1|do(X=0)) \nonumber \\
	&=& 14 \times \frac{14}{21} - 8 \times \frac{8}{21} = 9.33 - 3.04 = 6.29
\end{eqnarray}

where $\mathbf{1_d} \in \mathbb{R}^d$ is ones vector defined as $\mathbf{1_d} = (1, \dots, 1)^T$.

Another interesting thing to point out is the role subcortical region in the extracted network for the Bridgeville as a confounding node. In the connectivity matrix shown in figure \ref{fig:bridge_connectome}, the direction of edges are from rows to columns (causes to effects). This means that subcortical node drives 4 out of 6 nodes. This aligns with the fact that Bridgeville clip has emotional, motivational, musical and sport contents, which can stimulate subcortical regions like Amygdala. It also contains some old and nostalgic scenes, which can activate regions like hippocampus that is involved in processing of information related to memory. This is in accordance with the neuroscientific\- findings like \cite{tottenham2010review, richter2000amygdala} and emphasize on the key part that inclusion of subcortical regions have in the causal studies of the brain

\subsection{Future works}

In this study, we focused into one specific network in the brain, which decrease number of variables significantly. However, as stated previously, one of the problems in extracting a precise EC at the scale of whole-brain is related to discrepancy between the sizes of the whole-brain networks and available samples. We know that the edges in the EC are a subset of the edges in the structural connectome \cite{sporns2013structure}. So incorporating inforamtion of structural data as prior in favor of sparsity, can lead us to a more reliable result. Here, we investivgated the responce of face-selective regions to the movie watching data in a static framework. To gain more detailed insights, one can leverage dymanic methods in combination with HCP movie labels provided by \cite{huth2012continuous, nishimoto2011reconstructing}, and/or neural networks to extract the information of different frame and investigate how much do the results align.

\section{Conclusion}
In this paper, we investigated the causal structure of face-selective network in the brain in the naturalistic\- setting. we extracted causal networks of face-selective regions. The results showed that the presence of faces in the stimuli have causal effect on the number of identified causal connections within the network. Moreover, the results emphasize on the crucial role of subcortical regions in satisfying causal sufficiency, in causal studies of brain.

\newpage
\appendix
\setcounter{table}{0}  

\section{Coordinates of cluster centers}

\begin{table}[ht]
	\renewcommand{\arraystretch}{1.25}
	\caption{Coordinates of cluster centers on the flat map of cortex}
	\label{tab:coordinate}
	\centering
	\begin{tabular}{cc}
		\toprule
		Cluster & Center coordinate $(x,y)$ \\
		\midrule
		MFA\_L & $(137.783, 149.590)$ \\
		pSTS\_L & $(100.472, 10.090)$ \\
		FFA\_L & $(164.458, -18.507)$ \\
		MFA\_R & $(-125.207, 125.119)$ \\
		pSTS\_R & $(-107.803, -19.216)$ \\
		FFA\_R & $(-164.030, -44.811)$ \\
		\bottomrule
	\end{tabular}
\end{table}

\section{Metrics}
	\label{appendix}
Metrics used to assess synthetic data are defined as follows:
\begin{itemize}
	\item \textbf{Structural Hamming Distance (SHD):} A standard measurement for structure learning that counts the total number of edges additions, deletions, and reversals needed to convert the estimated graph into the true graph.
	
	\item \textbf{False Discovery Rate (FDR):} Measures the proportion of \textit{incorrectly} identified edges with respect\- to the total number of \textit{identified} edges.
	
	\item \textbf{False Positive Rate (FPR):} Measures the proportion of \textit{incorrectly} identified edges with respect to the total number of \textit{absent} edges in the ground-truth DAG.
	
	\item \textbf{True Positive Rate (TPR):} Measures the proportion of \textit{correctly} identified edges with respect\- to the total number of edges in the ground-truth DAG.
	
\end{itemize}

\newpage
\bibliographystyle{unsrtnat}
\bibliography{ref}
\end{document}